\documentclass[prl,aps]{revtex4}
\usepackage{epsfig}
\begin{document}
\draft
\widetext
\title{Detection of gravity waves by phase modulation of the light from a distant star}
\author{G. B. Lesovik$^a$, A. V. Lebedev$^{a,b}$, V. Mounutcharyan$^a$,
T. Martin$^{b}$}
\address{$^a$ Landau Institute for Theoretical Physics RAS, 117940 Moscow, Russia}
\address{$^b$ Centre de Physique Th\'eorique et  Universit\'e de la
M\'editerran\'ee, Case 907, 13288 Marseille, France}
\widetext

\begin{abstract}

We propose a novel method for detecting gravitational waves (GW),
where a light signal emitted from a distant star interacts with a
local (also distant) GW source and travels towards the Earth,
where it is detected. While traveling in the field of the GW, the
light acquires specific phase modulation (which we account in the
eikonal approximation). This phase modulation can be considered as
a coherent spreading of the given initial photons energy over a
set of satellite lines, spaced at the frequency of GW (from
quantum point of view it is multi-graviton absorption and emission
processes). This coherent state of photons with the energy
distributed among the set of equidistant lines, can be analyzed
and identified on Earth either by passing the signal through a
Fabry-Perot filter or by monitoring the intensity-intensity
correlations at different times.
\end{abstract}

\pacs{PACS }
\maketitle

\section{Introduction}
The detection of gravity waves (GW) has stimulated a lot of
interest for decades. There are two major GW detection concepts:
acoustic and interferometric detection. The acoustic method deals
with a resonance response of massive elastic bodies on GW
excitations. Historically the acoustic method was proposed first
by J. Weber~\cite{weber} where he suggested to use long and narrow
elastic cylinders as GW antennas. Although a significant progress
has been achieved in fabrication and increasing sensitivity of
such type of detectors~\cite{amaldi_1,amaldi_2,astone} the
interpretation of obtained data is still far to claim undoubtedly
the detection of GW. On the other hand a considerable attention
has been shifted recently to more promising interferometric
detection methods. The interferometric gravitational-wave detector
like Laser Interferometric Gravitational Wave Observatory (LIGO)
and VIRGO~\cite{vogt,bradaschia} represents a Michelson
interferometer with a laser beam split between two perpendicular
arms of interferometer. The principles of operation of such type
of detectors are reviewed in
Refs~\cite{forward,thorne,hough,drever,meers}. The action of
gravitational waves on an interferometer can be presented as
relative deformation of both interferometer arms. A gravitational
wave with dimensionless amplitude $h$ induces the opposite length
changes $\delta l/l=1/2 h\cos \Omega t$ in each arm of the
Michelson interferometer, where $l$ is the length of the arm,
$\Omega$ is the gravitational wave frequency. These length changes
produce opposite phase shifts between two light beams in
interferometer arms, when interference occurs at the beam splitter
of Michelson interferometer. The resulting phase shift of a single
beam of light spending time $\tau$ in the interferometer can be
written as~\cite{meers}
\begin{equation}
      \delta\phi=h\,\frac{\omega}{\Omega}\,\sin\frac{\Omega\tau}2,
      \label{laser_interferometer}
\end{equation}
where $\omega$ is the light frequency. This phase shift results an
intensity signal change of the light from interferometer beam
splitter hitting the photodetector.

The main problem of the acoustic and interferometric methods that
they both deal with gravitational waves with extremely small
amplitudes of the order $h\sim 10^{-21}$~\cite{schutz} reached the
Earth from deep space. One can see from
eq.~(\ref{laser_interferometer}) that for gravity wave frequencies
in the $1$ kHz range, $\Omega\sim 10^3$ Hz, and for the light in
visual frequency range, $\omega\sim 10^{14}$ Hz, one has the
maximum phase shift of the order $\delta\phi \sim 10^{-10}$ for
interferometer arms length of the order $150$ kilometers. Such
extraordinarily weak effect requires a exceedingly high detector
sensitivity both acoustic and interferometric detectors.

Alternatively, GW detection may be based on effects associated
with propagation of light or electromagnetic waves in
gravitational fields. There are two primary effects for the light
in constant gravitational field i) the deflection of light rays
near massive bodies~\cite{landau_2} and ii) the Shapiro effect
accounting for integrated time delay of the signal passing near
a strong source of gravitational field~\cite{shapiro}. The same
effects have to be observed for light propagating in gravity
waves: the gravitational waves have to induce  a weak time dependent
deflection of light ray propagating through these waves and also
lead to gravity-wave-induced variation in time delay. The idea to
use astrometry to detect periodic variation in apparent angular
separation of appropriate light sources was proposed by
Fakir~\cite{fakir_1}. It is shown that for a gravity wave source
located between the Earth and the light source (with line of sight
close aligned to gravity wave source) a periodical variation of
the order $\Delta\phi\sim \pi h(\Lambda)$ in the angular position of
the light source has to be observed. Here $h(\Lambda)$ is the
dimensionless strength of the gravity wave at distances of the
order gravitational wavelength $\Lambda$ which is many orders of
magnitude greater than the strength of the same waves when they
reach the Earth. On the other hand one can directly measure
the variation of the integrated time delay induced by gravity waves on the
light emitted by distant star which passing through space region
with strong gravity wave. The idea to use timing observation for
detection of the gravitational waves was suggested first by
Sazhin~\cite{sazhin_1} and then this problem was studied in
details by several authors~\cite{fakir_2,damour,kopeikin}. The
estimations carried out in Ref.~\cite{fakir_2} give the following
answer for the rate of change in the gravity-wave-induced time
delay, $\dot \tau$:
\begin{equation}
      |\dot \tau|\sim |h(r=D)|
      \label{time_delay}
\end{equation}
where $h(r=D)$ is the gravity wave strength at distances of
the order of impact parameter $D$ for the light beam passing near
the gravity wave source. To put some numbers let us consider a
dissymmetric rotating neutron star with spin frequency of the
order $10^3$ Hz and gravity wave amplitude of the order $H\sim
10^{-5}$ cm~\cite{thorne}, then eq.~(\ref{time_delay}) gives
$|\dot \tau|\sim 10^{-12}$ while the same effect measured near
Earth results $|\dot \tau|\sim 10^{-26}$ assuming that the neutron
star to be at a typical distance of a few kiloparsecs from the
Earth. One can see that all effects due to gravitational waves
near the gravity wave source are several orders of magnitude
stronger than the same ones on the Earth.

In the present paper we want to suggest a new gravitational waves
detection method based on the interaction of the photon with
gravitational waves. Assuming that the photons from distant star
passing near gravitational wave source, where the photon-gravity
wave interaction assumed to be strong, the photon-gravity wave
interaction leads to relatively strong modulation in time of
photon frequency. The latter allows to analyze this modulation of
the photons reaching the Earth by means of standard optical
methods including the Fabry-Perot analysis and quantum photon
correlations measurements. It is important that while the
interaction of photons with gravitational wave is rather weak
(proportional to strength of gravitational wave) the frequency
modulation can be accumulated over large distances during the
photon propagation that could result in an experimentally
measurable effect on the Earth. In some aspects our treatment of
the photon-gravity wave interaction resembles the effect of photon
acceleration by gravitational wave~\cite{mendonca}, where the
photon propagating in plane gravitational wave long enough time
acquires a considerable increase of the frequency. However while
in Ref.~\cite{mendonca} the photon-gravity wave interaction was
treated in the frame of reference of the photon, we have
considered this interaction in the point of observation frame
coordinate which seems to give rather the widening of the initial
monochromatic photon wave packet than the increase of the photon
frequency.

\section{Propagation of light near the localized source of gravitational waves.}

\begin{figure}
\centerline{\epsfxsize=7cm \epsfbox{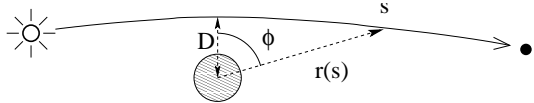}} \caption{A star
emits light which is deformed by the gravity field,
         but its phase is modulated, an effect which can be detected
         on the Earth.
\label{fig1}}
\end{figure}

In this section we consider how the plane electromagnetic waves
interact with gravitational wave field emitted by some localized
source. The situation we have in mind is depicted in Fig.
\ref{fig1}. The light signal originates from a distant star $S$
and travels towards the Earth to be detected by an observer $O$,
but along the way it interacts with a gravitational body $M$,
which also emits gravity waves. We consider the trajectory of the
light ray in the two dimensional plane formed by the star $S$, the
body which emits the GW, $M$, and the Earth. Provided that the
wave length of the GW is large compared to the wave length of the
light signal, we will use the eikonal or geometrical optics
approximation to describe the interaction of the light in
gravitational waves. The propagation of the electromagnetic waves
in the field of gravitational waves was considered many times and
we can refer the reader to several papers dealing with this
questions in the geometrical optics
approximation~\cite{cooperstock,estabrook,forward,braginsky,lobo,sazhin_2}.
Our subsequent analysis of this problem will be carried out in
close analogy with the paper of Sazhin and
Maslova~\cite{sazhin_2}, where they have considered the structure
of electromagnetic field in Fabry-Perot resonator in the field of
gravitational waves. However in the present paper we will be
interested in a different aspect of the interaction of the light beam
passing in immediate proximity to the gravitational wave source, where
the amplitude of the gravitational waves is assumed to be strong in
comparison with the amplitude of the same waves reaching the Earth.
It is important that since the light interacts with gravitational
waves in the region where they are strong, the resulting photons
frequency modulation seems to be appreciable to detect it on the
Earth.

\subsection{The eikonal equation.}

In the geometrical optics approximation the propagation of the light
ray is described by the eikonal equation:
\begin{equation}
      g^{ik}\,\frac{\partial\psi}{\partial x^i}\,
      \frac{\partial\psi}{\partial x^k}=0,
      \label{eikonal}
\end{equation}
where $g^{ik}=g^{ik}_0-h^{ik}$ the metric tensor associated with the
Schwarzschildian static metric $g^{ik}_0$ of the object $M$, which is
perturbed by the tensor $h^{ik}$ associated with the GW emitted from
$M$ \cite{landau_2}. Here we will neglect the true form of the static
metric $g^{ik}_0$ assuming that light ray propagates in the flat
Minkovsky space $g_0^{ik}=\eta^{ik}=\mbox{diag}\,\{1,-1,-1,-1\}$.
It is known that the static metric leads both to the deflection of
the light ray near the massive body and time delay of the light
signal passing along the ray (so called Shapiro effect). It can be
rigorously shown in the subsequent analysis that the former effect
of the light deviation leads to the negligible correction to the
additional phase accumulated by photons due to the interaction with GW
and one can assume the light trajectory to be a straight line. As
for static Shapiro time delay it can be easily incorporated to the
final answer for photon phase and actually does not affect on the
our subsequent detection method of modulated photons.

Let  $\psi_0=-\omega t+ky$ be the eikonal in the absence of gravitational
waves perturbation with metric $\eta^{ik}$ (we assumed the light ray
to be propagating along the $y$ axis in the plane formed by three bodies).
Assuming that the perturbation is small, in the presence of gravitational
waves the eikonal becomes $\psi=\psi_0+\psi_1$, where $\psi_1$ a small
addition to the eikonal $\psi_0$ computed to first order in $h^{ik}$
- it satisfies the equation:
\begin{equation}
      \frac{\partial\psi_1}{\partial t}+c\,\frac{\partial\psi_1}{\partial y}
      =-\frac\omega2\,F(t,r)\,,
      \label{first_order}
\end{equation}
where we have introduced the notation
$F(t,r)=h^{00}+h^{yy}-2h^{0y}$. Using the Green's function
formalism one can find the general solution of this equation.
Then to first order in the gravitational wave perturbation,
$h^{ik}$, the additional phase acquired by the photon due to
the interaction with the GW, can be written for arbitrary time dependence
of the perturbed metric:
\begin{equation}
      \psi_1(t,y)=-\frac12\,\frac\omega c\,\int\limits_{-\infty}^y
      F\left(t-\frac{y-y^\prime}c,y^\prime\right)\,dy^\prime
      \label{g_solution}
\end{equation}

\subsection{The emission of GW.}

Assuming that the GW tensor, $h^{ik}$, is a small perturbation to the
static flat metric in the quadrupole approximation one can write the
coordinate components of the GW tensor as~\cite{landau_2}:
\begin{equation}
      \tilde h_{\alpha\beta}=\frac{2G}{c^4}\,\frac{\ddot
      Q_{\alpha\beta}(t-r/c)}r
      \label{coordinate_GW}
\end{equation}
where $G$ is the Newtonian gravitation constant, $\tilde h_{ik}=h_{ik}-
\frac12\eta_{ik}h$ (where $h\equiv h_i^i$), $Q_{\alpha\beta}$ is the
quadrupole moment:
\begin{equation}
      Q_{\alpha\beta}(t)=\int \tilde T_{00}(t,r)\,x^\alpha x^\beta\,d^3x\,,
\end{equation}
where $\tilde T_{ik}=T_{ik}-\frac12\eta_{ik}T$, $T_{ik}$ is the
energy-momentum tensor of the GW source, $T\equiv T_i^i$.
Eq.~(\ref{coordinate_GW}) implies that the so called the harmonic
gauge condition, $\partial_k\tilde h_i^k=0$ has been chosen for
the components of GW tensor $h_{ik}$.

Using the gauge condition one can find the $\tilde h_{00}$ and
$\tilde h_{0\alpha}$ components of the GW tensor. To do this let us
write the equation $\partial_k\tilde h_i^k=0$ separately for
coordinate and time components:
\begin{equation}
      \frac{\partial \tilde h_{\alpha\beta}}{\partial x^\beta}=
      \frac{\partial\tilde h_{\alpha 0}}{\partial x^0}\,,\;\;\;\;\;
      \frac{\partial\tilde h_{0\alpha}}{\partial x^\alpha}=
      \frac{\partial\tilde h_{00}}{\partial x^0}\,.
      \label{gauge}
\end{equation}
Combining the eq.~(\ref{coordinate_GW}) with the first
eq.~(\ref{gauge}) the $\tilde h^{\alpha0}$ components of GW tensor
can be written as:
\begin{equation}
      \tilde h_{\alpha0}=\frac{2G}{c^3}\,\frac{\partial}{
      \partial x^\beta}\left[\frac{\dot Q_{\alpha\beta}(t-r/c)}r
      \right]\,,
      \label{coordinate_time_GW}
\end{equation}
while from this equation and the second equation~(\ref{gauge}) one
has:
\begin{equation}
      \tilde h_{00}=\frac{2G}{c^2}\,\frac{\partial^2}{\partial x^\alpha
      \partial x^\beta}\left[\frac{Q_{\alpha\beta}(t-r/c)}r\right]\,.
      \label{time_GW}
\end{equation}
Using the expressions~(\ref{coordinate_GW}),
(\ref{coordinate_time_GW}),(\ref{time_GW}) for the components of
the GW tensor write the explicit expression for
$F(t,r)=h^{00}+h^{yy}-2h^{0y}$ in the r. h. s. of
eq.~(\ref{g_solution}). Taking coordinate derivatives and keeping
in mind that along the ray trajectory $\partial_x r=D/r$,
$\partial_y r=y/r$, $\partial_z r=0$ (where $D$ is the impact
parameter) the $F(t,r)$ takes the form
\begin{eqnarray}
      F(t,r)&=&\frac{2G}{c^4}\left[\frac{\ddot Q_{xx}}r\,
      \frac{D^2}{r^2}+2\frac{\ddot Q_{xy}}r\,\frac Dr
      \left(\frac yr-1\right)+\frac{\ddot Q_{yy}}r\,
      \left(\frac yr-1\right)^2\right]
      \nonumber\\
      &+&\frac{2G}{c^3} \left[ 3 \frac{\dot
      Q_{xx}}{r^2}\,\frac{D^2}{r^2} + 2 \frac{\dot Q_{xy}}{r^2}\,
      \frac{D}{r} \left( 3\frac{y}{r} -1 \right) + \frac{\dot
      Q_{yy}}{r^2} \left( 3 \frac{y^2}{r^2}- 2 \frac{y}{r} -1
      \right)\right]
      \nonumber\\
      &+& \frac{2G}{c^2} \left[ 3 \frac{Q_{xx}}{r^3}\,
      \frac{D^2}{r^2} + 6 \frac{Q_{xy}}{r^3}\,\frac{yD}{r^2} +
      \frac{Q_{yy}}{r^3} \left( 3 \frac{y^2}{r^2} -1 \right)
      \right],
      \label{F}
\end{eqnarray}
where all quadrupole moments $Q_{\alpha\beta}$ are assumed to be
the functions of the retarded variable $t-r/c$.

\subsection{Solution of the eikonal equation.}

For simplicity the source of GW is assumed to emit only one
frequency $\Omega$. We consider the two following situations:
either the GW source has been emitting since $t=-\infty$, or the
signal is bounded in time, with an exponential decay. In the
former case one can write $\ddot
Q_{\alpha\beta}(t)=q_{\alpha\beta}\sin\Omega t$. Then substituting
the expression for $F(t,r)$, see eq.~(\ref{F}), into
eq.~(\ref{g_solution}) one could write the solution for $\psi_1$
as an integral over the variable $z = y^\prime/D -
\sqrt{1+(y^\prime/D)^2}$:
\begin{eqnarray}
      \psi_1(\xi)&=&-2\,\frac\omega c
      \int\limits_{D/(2y)}^\infty\frac{z^3 H_{yy}-2z^2 H_{xy}+
      z H_{xx}}{(z^2+1)^2}\sin\left[\Omega\xi-
      2\pi\frac D\Lambda z\right]dz
      \nonumber\\
      &+&4\frac{\omega}{c} \left( \frac{\Lambda}{2\pi D} \right)
      \int\limits_{D/(2y)}^\infty \frac{z^2(z^2-2)H_{yy}-2(2z^3-z)H_{xy}
      +3 z^2 H_{xx}}{(z^2+1)^3}
      \cos\left[ \Omega \xi - 2\pi \frac{D}{\Lambda} z\right] dz
      \nonumber\\
      &+& 4\frac{\omega}{c} \left( \frac{\Lambda}{2\pi D} \right)^2
      \int\limits_{D/(2y)}^\infty \frac{z(z^4-4z^2+1)H_{yy}
      -6z^2(z^2-1) H_{xy}+6z^3 H_{xx}}{(z^2+1)^4}
      \sin\left[ \Omega \xi - 2\pi \frac{D}{\Lambda} z\right] dz.
      \label{psi1_ex}
\end{eqnarray}
where the variable $\xi=t-y/c$ accounts for the effect of
retardation of the light signal during its propagation to the
point of observation, $H_{\alpha\beta}=2G q_{\alpha\beta}/c^4$,
$\Lambda$ is the wavelength of gravitational waves. Assuming that
at the point of observation $y\gg D$, one can safely set the lower
limit for all integrals in this expression equal to zero. Then for
large values of the impact parameter $2\pi D/\Lambda\gg1$ one can
asymptotically estimate these integrals and the main contribution
to $\psi_1$ is given by the following expression:
\begin{eqnarray}
      \psi_1(\xi)&\approx&\frac1{2\pi^2}\,\left(\frac\Lambda D\right)^2
      \,\frac\omega c\,H_{xx}\,\sin\Omega\xi
      \\
      &\approx&\frac{h_{xx}}\pi\,\frac\Lambda\lambda\,
      \left(\frac\Lambda D\right)^2\sin\Omega\xi\,,
      \label{eik_sol}
\end{eqnarray}
where in the last line the answer is rewritten in terms of the
dimensionless strength of gravitational waves:
$h_{xx}=H_{xx}\Lambda^{-1}$. One could see that the main
contribution to the photon phase $\psi$ comes from the $h_{xx}$
component of gravitational radiation confirming the well known
fact~\cite{damour,kopeikin} that only gravitational waves
propagating perpendicular to the photon wave vector (directed
along $y$ axis in our geometry) have a considerable effect on the
electromagnetic radiation.

Let us now consider the case where gravity waves are emitted
starting at $t=0$ decaying exponentially with characteristic time
$\tau_0$. In this case let us choose $\ddot Q_{\alpha\beta}(t)$ to
be equal to:
\begin{equation}
      \ddot Q_{\alpha\beta}(t)=q_{\alpha\beta}\,\theta(t)
      \,\exp\left(-\frac{t}{\tau_0}\right)\sin\Omega t\,,
      \label{bounded}
\end{equation}
where $\theta(x)=0$ for $x<0$ and $1$ for $x>0$. Then for large
values of the impact parameter, $2\pi D/\Lambda\gg1$, the solution for
$\psi_1$ is given by the following formula:
\begin{eqnarray}
      \psi_1(\xi)&\approx&\frac{h_{xx}}{\pi}\left(\frac\Lambda D\right)^2
      \frac\Lambda\lambda \,\,\theta(\xi)\,
      e^{-\xi/\tau_0}\,\sin\Omega\xi\,.
\end{eqnarray}
One can see that in this case at the point of observation the
modulated contribution to the eikonal also decaying exponentially
with the same characteristic time $\tau_0$.

\subsection{The pre-exponential factor.}

Let us now go beyond the geometrical optics or eikonal
approximation and calculate the pre-exponential factor for the
photon wave function. The photon wave function $\varphi(t,y)$ with
momentum $k$ satisfies the wave equation:
\begin{equation}
      \frac1{\sqrt{-g}}\,\frac{\partial}{\partial x^i}\left(
      \sqrt{-g} g^{ik}\frac{\partial \varphi}{\partial x^k}
      \right) = 0,
      \label{cov_we}
\end{equation}
where $g=\det g_{ik}$. Up to the first order contribution in
$h^{ik}$ one could find $g=-(1+h)$ with $h= h_i^i$. Taking the
derivative over $x^i$ and keeping only the first order terms in
$h^{ik}$ one could arrive to expression:
\begin{equation}
      g^{ik}\frac{\partial^2 \varphi}{\partial x^i \partial x^k} -
      \frac{\partial \varphi}{\partial x^k}
      \frac{\partial}{\partial x^i} \left( h^{ik} -\frac12 \eta^{ik}
      h \right) =0.
\end{equation}
One could see that under the harmonic gauge condition $\partial_i
\tilde h_k^i=0$ with $\tilde h_k^i= h_k^i -\frac12 \eta_k^i h$ the
second term in above equation vanishes and the photon wave
equation reduces to
\begin{equation}
      g^{ik}\frac{\partial^2 \varphi}{\partial x^i \partial x^k} =
      0.
\end{equation}

Let us write the photon wave function as
$\varphi(t,y)=c(t,y)e^{i\psi}$, where $\psi$ is the eikonal of the
photon interacting with gravitational waves. Then assuming that
$c=1+a$, where $a$ is the small contribution of the first order in
$h^{ik}$, one could rewrite the above equation as
\begin{equation}
      g^{ik}\left( \frac{\partial^2 a}{\partial x^i \partial x^k}
      +2i \frac{\partial \psi}{\partial x^i} \frac{\partial
      a}{\partial x^k} + i(1+a) \frac{\partial^2 \psi}{\partial x^i
      \partial x^k} - (1+a) \frac{\partial \psi}{\partial x^i}
      \frac{\partial \psi}{\partial x^k}
      \right)=0.
\end{equation}
The last term in this equation vanishes due to the eikonal
equation~(\ref{eikonal}). Since the free eikonal $\psi_0=ky-\omega
t$ is the linear function of $x^i$ the all second derivatives of
$\psi_0$ are vanish and the remaining wave equation up to the
first order terms in $h^{ik}$ takes the form:
\begin{equation}
      \eta^{ik} \frac{\partial^2 a}{\partial x^i \partial x^k} +
      2i\eta^{ik} \frac{\partial \psi_0}{\partial x^i}
      \frac{\partial a}{\partial x^k}+i\eta^{ik}\frac{\partial^2
      \psi_1}{\partial x^i \partial x^k}=0.
\end{equation}
Assuming that the $a(t,y)= a(t-y/c)$ thus the first term above
vanishes one could finally find
\begin{equation}
      \frac{\partial a}{\partial t}+c\frac{\partial a}{\partial y}=
      \frac12\,\frac{c^2}\omega\left(\frac1{c^2}\frac{\partial^2\psi_1}{
      \partial t^2}-\frac{\partial^2\psi_1}{\partial y^2}\right).
      \label{pre_g}
\end{equation}
Substituting here the general solution for $\psi_1$ (see eq.~(\ref{g_solution}))
one can express the r. h. s. of eq.~(\ref{pre_g}) through the function $F(t,y)$:
\begin{equation}
      \frac{\partial a}{\partial t}+c\frac{\partial a}{\partial y}=
      \frac c4\left(\frac{\partial F}{\partial y}-\frac1c\frac{
      \partial F}{\partial t}\right)
      \label{pre_p}
\end{equation}
According to our previous analysis the main contribution to $a(t,y)$ is
given by the $H_{xx}$ component of gravitational radiation:
\begin{equation}
      F(t,r)\approx H_{xx}\frac{\sin\Omega(t-r/c)}r\,\frac{D^2}{r^2}
\end{equation}
Substituting this expression to the eq.~(\ref{pre_p}) the solution
for $a(\xi)$ for large values of impact parameter $2\pi
D/\Lambda\gg1$ has the form:
\begin{equation}
      a(\xi)\approx -\frac{h_{xx}}{\pi}\,\left(\frac\Lambda D\right)^2
      \cos\Omega\xi
      \label{preexp_sol}
\end{equation}
Combining together eq.~(\ref{eik_sol}) with eq.~(\ref{preexp_sol}) one
can finally write the photon wave function at the point of observation
in the form:
\begin{equation}
       \varphi_k(\xi)=\frac{e^{-i\omega(\xi+\tau_g\sin\Omega\xi)}}{
       1+\tau_g\Omega\cos\Omega\xi}\,,
       \label{photon_func}
\end{equation}
where we have introduced the time $\tau_g$ associated with
integrated Shapiro time delay (see below):
\begin{equation}
      \tau_g=-\,\frac{h_{xx}}{\pi}\left(\frac\Lambda D\right)^2\,
      \Omega^{-1}\,.
      \label{delay}
\end{equation}
The same solution eq.~(\ref{photon_func}) for the photon wave function is
valid for the case where gravitational radiation are bounded exponentially
in time (see eq.~(\ref{bounded})) however in this case $\tau_g$ depends on
time and equals to
\begin{equation}
      \tau_g(\xi)=-\frac{h_{xx}}{\pi}\,\left(\frac\Lambda D\right)^2\,
      \Omega^{-1}\,\theta(\xi)\,e^{-\xi/\tau_0}\,.
\end{equation}

\subsection{Modulation due to alternating Newtonian gravitational field.}

The observation and detection of the gravitational waves undoubtedly
provides us a new astrophysical tool which can give us a deeper insight
on the processes like the neutron star formation, the processes
happening in close binaries at the final stage of it evolution, etc....
However the observation of gravitational waves gives us also an additional
proof of the General Theory of Relativity itself and it seems
reasonable to consider the effect of photon phase modulation in
Newtonian theory of gravitation where the gravitational potential
instantly adjusts to the current configuration of moving bodies.

Consider for example an oscillating neutron star or a rotating
double star system which in Newtonian theory of gravity causes an
alternating Newtonian gravitational potential $\phi({\bf r},t)$.
This alternating field can also lead to photon phase modulation
and results in principle in the same effect like for the case of
photon in the field of gravitational waves. The aim of this
section is to compare the effect of photon phase modulation
between these two theories.

As the rotation or the oscillation frequency of the stars  is
much less than the frequency of the photon we can work in the same
eikonal approximation as in the previous sections. The Newtonian
gravitational field contributes only to the diagonal metric
elements: $h^{00}=h^{ii}=2\phi/c^2$, where $\phi$ - is a
gravitational potential at the point of observation. At large
distances compared to the size of the star system we can
leave only quadrupole term in the potential
\begin{equation}
      \phi({\bf r},t) =
      G Q_{\alpha \beta}\, \frac{n_{\alpha} n_{\beta}}{2 r^3}.
\end{equation}
Here $r$ - is the distance from the star to the photon, the
coordinate axis are chosen to be the same as before, so $r=
\sqrt{y^2 + D^2 }$, $n_{i}$ - the direction vectors of ${\bf r}$:
$n_x=D/\sqrt{y^2+D^2}$, $n_y = y/\sqrt{D^2+y^2}$, $n_z=0$,
$Q_{\alpha \beta}(t)$ is the alternating tensor of the mass
quadrupole moment of the star. The values of $Q_{\alpha \beta}$
are determined by the parameters of the star, like orientation of
the angular velocity, eccentricity, etc. We will further discuss
the general case, assuming only, that due to rotation or
oscillation $\ddot Q_{\alpha \beta}(t) = q_{\alpha
\beta}\sin\Omega t$.

The correction to the eikonal due to rotation is given by the same
equation~(\ref{g_solution}) as for the GW case solution of the
eikonal equation, where in our case $F(t,y)=h^{00}+h^{yy}=2h^{00}$
\begin{equation}
      \Psi_n = -\frac\omega{c}\int\limits_{-\infty}^{y}h^{00}
      \left(t-\frac{y-y^\prime}c,y^\prime\right)\,dy^\prime
      \label{eik_newt}
\end{equation}
Substituting the expression for $\phi$ into the formula above and
making the substitution $z=y^\prime/D$ one obtains:
\begin{eqnarray}
       \Psi_n(\xi)&=&\frac12 \frac\omega{c}
       \left(\frac{\Lambda}{2\pi D}\right)^2
       \int\limits_{-\infty}^{y/D}\frac{H_{xx}+2H_{xy}z+H_{yy}z^2}{
       (1+z^2)^{5/2}}\sin\left[\Omega\xi+
       2\pi\frac{D}\Lambda z\right] dz,
\end{eqnarray}
where we have introduced $H_{\alpha\beta}=2Gq_{\alpha\beta}/c^4$.

Then for large values of the impact parameter $2\pi D/\Lambda\gg1$
and large distances $y/D\gg1$ one can obtain the following approximate
expression for $\Psi_n$:
\begin{eqnarray}
      \Psi_n(\xi)=\frac{1}{6} \frac{\omega}{c}\,\sqrt\frac\Lambda{D}\,
      e^{-2\pi D/\Lambda}\,\bigl[(H_{xx}-H_{yy})\sin\Omega\xi
      +2H_{xy}\cos\Omega\xi \bigr].
\end{eqnarray}

One can see that the photon phase modulation due to the Newtonian
alternating potential vanishes exponentially for large impact
parameters in contrast to the GW modulation which vanishes as
$1/D^2$ (see eq.~(\ref{eik_sol})). Technically the difference
between Newtonian and GW cases follows from the different
$y^\prime$ dependence of the $F(t,y^\prime)$ function. For the
properly retarded GW radiation the solution for $\psi_1$, see
eq.~(\ref{eik_sol}), involves the oscillating integral over
$\sin\Omega[\xi +(y^\prime -\sqrt{(y^\prime)^2+D^2})/c]$, where
the oscillating dependence on $y^\prime$ near the source
$|y^\prime|\sim D$ almost vanishes making the light-GW interaction
more effective. Oppositely for the instantly adjusted Newtonian
radiation one has in eq.~(\ref{eik_newt}) fast oscillating
integral over $y^\prime$ with $\sin \Omega(\xi+y^\prime/c)$ near
vicinity of the gravitational wave source, which leads to a very
small value of modulation.

\subsection{The Shapiro effect in GW.}

Using the results of the previous sections let us calculate the
integrated time delay of the light signal propagating in the field
of gravitational waves. Consider an arbitrary wave packet emitted
by some distant light source traveling toward to Earth near a
source of GW:
\begin{equation}
      \varphi(\xi)=\int \varphi_\omega\,e^{-i\omega\xi}\,
      \frac{d\omega}{2\pi}
\end{equation}
where $\xi=t-y/c$. Using the results for monochromatic, photons propagating
in the field of gravitational radiation one can find the signal at the point
of observation:
\begin{eqnarray}
      \varphi_{ob}(\xi)&=&\int \varphi_\omega\,
      e^{-i\omega[\xi+\tau_g\sin\Omega\xi]}\,\frac{d\omega}{2\pi}
      \nonumber\\
      &=&\varphi(\xi+\tau_g\sin\Omega\xi).
\end{eqnarray}
One can see that the interaction with the gravitational radiation
leads to alternating time delay of the signal $\delta
t=\tau_g\sin\Omega t$, (note that the $\tau_g$ is negative, see
eq.~(\ref{delay})). This effect was suggested independently first
by Sazhin and Detweiler~\cite{sazhin_1} for detection of
gravitational waves using the timing observation of a pulsar the
line of sight to which passes near the source of GW.

\section{Detection of the GW}

Let us now turn to the discussion about how the photons modulated by
the interaction with the gravitational waves can be measured in some
realistic setup.

Consider first the simple case where the modulated
photons~(\ref{photon_func}) simply hit a photodetector which can
react at all photon's frequencies. However, any real photodetector
reacts not on the vector potential ${\bf A}(t)$ of the photons
in the photodetector directly, but rather on the electrical field,
${\bf E}(t)$, induced by the photons in this photodetector. Then
the one photon photo-detection probability, $P(t)\Delta t$, to
observe a photon during time interval $\Delta t$ can be
expressed as~\cite{mandel}:
\begin{equation}
      P(t)\Delta t=\alpha\,\frac{Sc\Delta t}{E_{det}}
      \left\langle{\bf\hat E^{(-)}}(t,{\bf r})
      \cdot{\bf\hat E^{(+)}}(t,{\bf r})\right\rangle\,
\end{equation}
where $S$ is the area of the photodetector, $E_{dec}$ some
characteristic energy describing the interaction of the
photodetector with incoming photons, $\alpha$ is the quantum
efficiency of the photo-detection, ${\bf\hat E^{(+)}}(t,{\bf r})$
is the positive frequency part of the electrical field operator,
${\bf\hat E^{(-)}}(t,{\bf r})$ is the Hermitian conjugate of
${\bf\hat E^{(+)}}(t,{\bf r})$:
\begin{equation}
      {\bf\hat E^{(+)}}(t,{\bf r})=-\,\frac{\sqrt{4\pi}}{L^{3/2}}
      \,\sum\limits_{{\bf k}s}\sqrt\frac\hbar{2\omega_k}\,\,
      \dot\varphi_k(t,{\bf r})\,{\bf e}_{ks}\,\hat a_{ks}\,.
      \label{pos_e_op}
\end{equation}
At the point of observation we set ${\bf r}=0$ to shorten the
notations. Taking the time derivative of the photon wave
function~(\ref{photon_func}) one gets the photo-detection
probability equals to
\begin{eqnarray}
      P(t)\Delta t&=&\frac{4\pi\alpha\,Sc\Delta t}{E_{det}L^3}
      \sum\limits_{{\bf k}s}\frac{\hbar\omega_k}2\,n_{ks}
      \nonumber\\
      &=&2\alpha\,\frac{S}{r^2}\,\Delta t
      \int d\omega\,\frac{\hbar\omega}{E_{det}}\,n(\omega)\,,
\end{eqnarray}
where $r$ is the distance between the light source and the Earth,
$n(\omega)\,d\omega$ is the number of the photons with frequency
$\omega$, emitted by the distant light source per unit time, into
a unit angular domain.

One can see that the photo-detection probability does not feel
any photon phase modulation due to GW and this probability
is the same as it would be for the monochromatic photons emitted
by a distant light source. Thus for such a simple setup it is not
possible to retrieve any information about the gravity wave from
the photodetector signal.

Consider however a more realistic setup where the photodetector
has a sensitivity edge $\omega_s$ which means that photodetector
does not react on the photons with frequency less than $\omega_s$.
One can treat this situation as if the photons passed through a
filter with transparency $t(\omega)=\theta(\omega-\omega_s)$,
before hitting the photodetector. It is convenient to write this
transparency as an operator acting on the photon wave function:
$\hat t=\theta[i\partial_t-\omega_s]$. Then after passing such a
filter the photon has a wave function equals to $\hat
t\varphi_k(t)\approx\theta[\omega_k(t)-\omega_s] \varphi_k(t)$,
where $\omega_k(t)=\omega_k[1+\Omega\tau_g\cos\Omega t]=\omega_k
f(t)$. Then for the photo-detection probability one has:
\begin{eqnarray}
      P(t)\Delta t\approx 2\alpha\,\frac{S\Delta t}{r^2}
      \int d\omega \frac{\hbar\omega}{E_{det}}\,\,
      \theta[\omega f(t)-\omega_s]\,n(\omega).
      \label{edge_sig}
\end{eqnarray}
Consider now the case where photons coming to the photodetector
from a sharp Lorentzian spectral line with maximum, $\omega_0$,
near the photodetector sensitivity edge $\omega_0\approx\omega_s$
and width $\Gamma_0$:
\begin{equation}
      n(\omega)=n(\omega_0)\frac{\Gamma_0^2}{(\omega-\omega_0)^2+
      \Gamma_0^2},
      \label{spectral_line}
\end{equation}
then according to the eq.~(\ref{edge_sig}) the photo-detection
probability for  the modulated photons~(\ref{photon_func}) has
an appreciable contribution which is periodic in time
at the GW frequency, and which is
proportional to the strength of gravitational waves:
\begin{equation}
      P(t)\Delta t\approx P_0\Delta t
      \left(1+\Omega\tau_g\frac{2\omega_0}{\pi\Gamma_0}
      \,\cos\Omega t\right),
\end{equation}
where $P_0\Delta t$ is the photo-detection probability for non
modulated photons.

From the above analysis one can conclude that in order to detect the
modulation of the photons by GW one should place before the
photodetector some filter which has a finite frequency band. Then
the formula for photo-detection probability~(\ref{edge_sig}) with a
sensitivity edge has a rather general sense. In fact let us
consider an arbitrary filter placed before the photodetector with
transparency $T(\omega)$. Then provided that the time spent by the
photon in this filter (that is an inverse frequency band of the
filter) is much smaller than frequency of the GW, $\Omega$, one can
treat the modulated photons as monochromatic photons with slowly
varying frequency $\omega_k(t)=\omega_k[1+\Omega \tau_g\cos\Omega
t]=\omega_k f(t)$. In this case the photo-detection probability can
be written simply as
\begin{equation}
      P(t)\Delta t\approx 2\alpha\,\frac{S\Delta t}{r^2}
      \int d\omega\frac{\hbar\omega}{E_{det}}\, T(\omega f(t))\,
      n(\omega)\,.
      \label{simple}
\end{equation}

From this expression one can see that there are two different
regimes of GW detection in this proposed setup. Consider the first
regime where the distribution function of the light source,
$n(\omega)$, has a sharp peak of the width $\Gamma_0$ (or in more
general case some sufficient irregularity like the edge of the
spectrum) centered near the transmission window of the filter.
Then, provided that the amplitude of the photon's frequency
modulation, $\omega_k\Omega\tau_g$, is much bigger than the width of
the spectrum irregularity, $\omega_k\Omega\tau_g\gg\Gamma_0$, the
resulting photo-detection signal will be a periodic sequence of
peaks.

In the opposite case where the light spectrum $n(\omega)$ is a
slowly varying function of the frequency within the transparency
window of the filter one can consider $n(\omega)$ as a constant in
eq.~(\ref{simple}). Then the time dependence of the photo-detection
probability can be written as:
\begin{equation}
      P(t)\Delta t\approx 2\alpha\,\frac{S\Delta t}{r^2}\,
      \frac1{f(t)}
      \int d\omega\frac{\hbar\omega}{E_{det}}\, T(\omega)\,
      n(\omega)\,.
\end{equation}
This result can be understood as the time modulation of the number
of incoming photons per unit time within a finite frequency
interval.

From another point of view one can note that the modulated photon
wave function~(\ref{photon_func}) can be written as a superposition of
waves with energies shifted by $n\Omega$ ($n$ integer):
\begin{equation}
       \varphi_{k}(t)=\frac1{1+\Omega\tau_g\cos\Omega t}\sum_{-\infty}^{+\infty} J_n(\omega \tau_g)
       \,e^{-i(\omega+n\Omega)t}
       \label{time_fourier}
\end{equation}
where $J_n$ is the Bessel function. In order to detect a gravity wave
signal it is therefore necessary to provoke an interference between
these Fourier components. To do it one can both investigate the modulated
light signal passing through the Fabry-Perot interferometer or from
the other hand study the intensity-intensity correlation of the photons
coming on the photodetector.

\subsection{Analysis of the light signal with an interferometer}

In this section we consider in detail the setup where at the point
of observation - the Earth - the modulated light  signal before hitting
the photodetector passes through an interferometer, which is characterized
by a complex transmission amplitude $t(\omega)$. We will assume that
the light passing through a Fabry-Perot filter comes from a single Lorentzian
spectral line of the type~(\ref{spectral_line}). For simplicity, let us
first assume that there is only one FP resonance within this spectral line.
The transparency of FP is written in the usual way:
\begin{equation}
      t(\omega)=\frac{i\Gamma/2}{\omega-\omega_0+i\Gamma/2}.
\end{equation}
Here a derivation of the photo-detection probability is provided.
It relies on the assumption $\Omega\tau_g\ll1$, which is relevant
for experimental situations (see below).

Consider the propagation of a photon wave-packet
$\varphi_{k}(t)$~(\ref{photon_func}) through the FP
interferometer. The resulting wave packet after the FP can be
written $\phi_k(t)=\hat t\,\varphi_k(t)$, where $\hat t$ is the
transparency operator of the FP in real time representation:
\begin{equation}
      \hat t =\frac{i\Gamma}2\,[i\partial_t-\omega_0+i\Gamma/2]^{-1}\,,
\end{equation}
where $\Gamma$ is the width of FP resonance. However the real FP
filter does not act directly on the vector potential of the
incoming photons but rather on the electromagnetic field induced
by the these photons. Using similar arguments the positive
frequency part of electrical field operator ${\bf E}^{(+)}(t,r)$,
after passing the Fabry-Perot filter, can be written in terms of
the transmission operator of the filter:
\begin{equation}
      {\bf\hat E^{(+)}}(t,{\bf r})=-\,\frac{\sqrt{4\pi}}{L^{3/2}}
      \,\sum\limits_{{\bf k}s}\sqrt\frac\hbar{2\omega_k}\,\,
      \bigl[\hat t\dot\varphi_k(t,{\bf r})\bigr]\,{\bf e}_{ks}\,\hat a_{ks}\,.
\end{equation}

With these notations the photo-detection probability can be written
as
\begin{equation}
      P(t)=2\alpha\frac{S}{r^2}\int \frac{\hbar d\omega}{\omega E_{det}}\,
      \bigl[\hat t\dot\varphi_k(t)\bigr]^*\bigl[\hat
      t\dot\varphi_k(t)\bigr]\,n(\omega)\,d\omega.
      \label{prob_sing_g}
\end{equation}
The action of the operator $\hat t$ on the function $\dot \varphi_k(t)$
can be found using the Green's function formalism:
\begin{equation}
      \hat t\dot\varphi_k(t)=\frac{\Gamma}2
      \int\limits_{-\infty}^t e^{-i\omega_0(t-\tau)-\frac\Gamma2(t-\tau)}
      \dot \varphi_k(\tau)\,d\tau.
\end{equation}
Substituting this expression into eq.~(\ref{prob_sing_g}) in the
limit $\omega_0\gg\Gamma$ one finally obtains the following
expression for the photo-detection probability:
\begin{eqnarray}
      P(t)=2\alpha\frac{S}{r^2}\,\frac{\hbar\omega_0}{E_{det}}
      \frac{\Gamma^2}4\,\int n(\omega)
      \,e^{-\Gamma t}d\omega
      \int\limits_{-\infty}^t\int\limits_{-\infty}^t e^{-i\omega_0(\tau-s)
      +\frac\Gamma2(\tau+s)}\dot \varphi_k^*(\tau)\dot \varphi_k(s)
      \,d\tau d s
\end{eqnarray}
Performing the integral over $\omega$ and introducing the relative
time $\xi=\tau-s$, and the total time $\eta=(\tau+s)/2$, in the
working limit $\Omega\tau_g\ll1$ one can write the photo-detection
probability as:
\begin{eqnarray}
     P(t)=\alpha\pi\,\frac{S}{r^2}\,\frac{\hbar\omega_0}{E_{det}}
     \,n(\omega_0)\Gamma_0\,\frac{\Gamma^2}4
     \int\limits_{-\infty}^t
     d\eta\,e^{-\Gamma(t-\eta)}
     \int\limits_{-2(t-\eta)}^{2(t-\eta)}
     e^{i\xi[\omega_0\Omega\tau_g
     \cos\Omega\eta]}\,e^{-\frac{\Gamma_0(\eta)}2|\xi|}\,d\xi
\end{eqnarray}
Performing the integration over $\xi$ in the limit
$\Gamma_0\gg\Gamma$ the leading contribution to the photo-detection
probability is given by the following expression:
\begin{eqnarray}
      P(t)=\alpha\pi\,\frac{S}{r^2}\,\frac{\hbar\omega_0}{E_{det}}\,
      n(\omega_0)\Gamma_0\,\frac{\Gamma^2}4
      \int\limits_0^\infty\frac{\Gamma_0(t-z)\,e^{-\Gamma z}\,dz}{
      (\omega_0\Omega\tau_g)^2\cos^2
      \Omega(t-z)+\Gamma_0^2(t-z)/4}
      \label{strong_ex}
\end{eqnarray}
where we have made the substitution $z=(t-\eta)$ and we introduced the
notation $\Gamma_0(t)=\Gamma_0|1+\Omega\tau_g\cos\Omega t|$.

\subsubsection{''Strong'' GW or narrow spectrum.}

Let us consider the limit where the frequency broadening of the
initially monochromatic photon wave packet due to the interaction with
gravitational waves, $\delta\Gamma=\omega_0\Omega\tau_g$, is much
bigger than the width of the spectral line, $\Gamma_0$. The
condition $\delta\Gamma\gg\Gamma_0$ can be achieved both in the
case where the amplitude of GW is strong enough or when one has a
very narrow spectral line. In this limit one can safely neglect
the dependence on $|1+\Omega\tau_g\cos\Omega t|$ and put
$\Gamma_0(t)=\Gamma_0$ in eq.~(\ref{strong_ex}) assuming
$\Omega\tau_g\ll1$. Then the leading contribution to the
photo-detection probability can be written in terms of the
occupation number, $n(\omega)$, of the spectral line:
\begin{equation}
      P(t)=\alpha\pi\,\frac{S}{r^2}\,\frac{\hbar\omega_0}{E_{det}}
      \;\Gamma^2\int\limits_0^\infty
      n[\omega_0(t-z)]\,e^{-\Gamma z}\,dz
      \label{strong}
\end{equation}
where $\omega_0(t)=\omega_0(1+\Omega\tau_g\cos\Omega t)$. This result has
a very simple qualitative explanation which is consistent with our previous
simple arguments for the photodetector with sensitivity edge (see eq.~(\ref{simple})).
Consider a photon wave packet coming to FP filter: $\varphi_k={\cal C}(t)
e^{-i\omega_k(t+\tau_g\sin\Omega t)}$. Such a wave packet in the limit
$\Omega\ll\omega_k$ can be interpreted as a photon with a frequency which is
slowly varying
in time  $\omega_k(t)=\omega_k(1+\Omega\tau_g\cos\Omega t)$. If
however the time spent by this photon in the FP filter, $\Gamma^{-1}$, much
smaller than the characteristic time of the frequency change $\Omega^{-1}$ one
can assume the photon at any instant of time to be monochromatic with frequency
$\omega_k(t)$. Then one can take advantage of our eq.~(\ref{simple}).
Performing the integration over $\omega$ in eq.~(\ref{simple}) with
Lorentzian spectral line $n(\omega)$~(\ref{spectral_line}) one has
\begin{equation}
      P(t)=\alpha\pi\,\frac{S}{r^2}\,\frac{\hbar\omega_0}{E_{det}}\,
      n[\omega_0(t)]\,\Gamma.
      \label{quick}
\end{equation}
On the other hand this result can be derived directly from our
elaborate analysis by putting $n(\omega_0(t-z))=n(\omega_0(t))$ in
eq.~(\ref{strong}). The exponential factor in eq.~(\ref{strong})
describes a time delay of the photon wave packet propagating
through the FP filter.

The result~eq.(\ref{quick}) describes the periodic sequence of symmetric
peaks with a half period of gravitational waves $\pi/\Omega$ and widths equal to
\begin{equation}
      \tau=\frac{\Gamma_0}{\delta\Gamma}\,\Omega^{-1}.
      \label{width}
\end{equation}
If however the width of these peaks becomes smaller than the time
needed for the photon to penetrate through the FP filter:
$\tau\ll\Gamma^{-1}$ then the form of peaks becomes asymmetric and
the more accurate expression~(\ref{strong}) has to be used.

It should be noted that the expression~(\ref{strong}) is valid not
only for a single spectral line~(\ref{spectral_line}) but also for
arbitrary distribution function for the light source, where there
is a sufficient irregularity in $n(\omega)$ like, for example, at
the edge of the spectrum. Indeed if one consider the arbitrary
distribution function $n(\omega)$ which changes much at frequency
scale $\Delta\omega$ around $\omega_0$ then it follows from
eq.~(\ref{strong}) that one again has a periodical sequence of
peaks for photo-detection signal in regime $\delta\Gamma\gg\Delta
\omega$. In this sense the described effects~(\ref{strong}),
(\ref{quick}) have nothing to do with the interference of the
different components of the photon wave
packet~(\ref{time_fourier}) and deal rather with the effect of the
photon frequency modulation $\omega_k(t)=\omega_k(1+\Omega
\tau_g\cos\Omega t)$ due to the interaction with the GW. If
however the amplitude of the frequency modulation $\delta\Gamma$
becomes smaller than the characteristic scale of the spectral
function $n(\omega)$ (the case of a broad spectrum) the effects
associated with frequency modulation completely disappear.

\subsection{''Weak'' GW or broad spectrum.}

Consider now the opposite limit to the previous case when the
amplitude of GW is small or when the width of the spectral line is
large enough so that the amplitude of the frequency modulation,
$\delta\Gamma$,  is much smaller than the width of the spectral line,
$\Gamma_0$. It should be noted that this situation is the most
common situation in practice since the amplitude of GW is
extremely small even in the vicinity of GW source. In the limit
$\delta\Gamma\ll\Gamma_0$ one can safely expand the integral
expression in eq.~(\ref{strong_ex}) up to the lowest order in the
small ratio $\delta\Gamma/\Gamma_0\ll1$ and perform the
integration over $z$. Keeping all terms of the type
$1+\Omega\tau_g \cos\Omega t$ in the working assumption
$\Gamma\gg\Omega$ the photo-detection probability can be now
written as:
\begin{eqnarray}
      P(t)\approx\alpha\pi\,\frac{S}{r^2}\,\frac{\hbar\omega_0}{E_{det}}\,
      n(\omega_0)\Gamma\,
      \left(1-\Omega\tau_g\cos\Omega t
      -2\frac{\delta\Gamma^2}{
      \Gamma_0^2}\cos2\Omega t\right)
      \label{weak}
\end{eqnarray}
One can see that in the case of a broad spectrum one obtains small
oscillations against a huge constant background with frequencies
$\Omega$ and $2\Omega$ suppressed by small factors
$\Omega\tau_g\ll1$ and $\delta\Gamma^2/\Gamma_0^2\ll1$,
correspondingly. However as one can see from eq.~(\ref{weak}) even
for the extreme case of infinitely broad spectral line still there is
a time dependent contribution to the $P(t)$:
\begin{equation}
      P(t)\approx\alpha\pi\,\frac{S}{r^2}\,\frac{\hbar\omega_0}{E_{det}}\,
      n(\omega_0)\Gamma\,\bigl[1-\Omega\tau_g\cos\Omega t\bigr]
      \label{weak_b}
\end{equation}
It should be noted that this expression does not depend absolutely on
the form of the spectra of the distant light source and the time modulation
of the photo-detection signal defined only by the strength of the
gravitational waves. The amplitude of the photo-detection probability
in this case proportional to the number of photons $n(\omega_0)\Gamma$
coming to the photodetector per unit time and practically one can use
the FP filter with appropriate width $\Gamma$ to collect the enough
number of photons.

The second important point which has to be emphasized here is that
the time dependent term in eq.~(\ref{weak_b}) has nothing to do
with the effect of the frequency modulation mentioned above (see
eq.~\ref{strong}, \ref{quick}) and presents an absolutely
different effect associated with interference of the different
components of the photon's wave packet~(\ref{time_fourier}) after
passing the FP filter. To clarify the nature of this interference
effect we will consider in the next section an example where there
are two extremely narrow FP resonances, $\Gamma\ll \Omega$, within
the broad spectrum of the distant light source.

\subsection{Two resonances case.}

Let us consider a more general case where two narrow resonances at
frequencies $\omega_0\pm\Delta/2 $ are present within the spectral
line ($\Delta$ is the separation between resonances). In this case
the transmission amplitude of the FP filter, $t(\omega)$ again can be
written as an operator acting on the e.m.f. induced by modulated
photons: $\hat t= \hat t_1+\hat t_2$, where $\hat t_{1(2)}$
corresponds to the first (second) resonance:
\begin{equation}
      \hat t_{1(2)}=e^{\pm i\phi}\frac{i\Gamma}2
      \left[i\partial_t-\omega_0\pm\frac\Delta2+i\frac\Gamma2
      \right]^{-1},
\end{equation}
where $2\phi$ is the relative phase difference between resonances.

Using similar type of arguments and the same method of
calculation as for the case of FP filter with a single resonance,
in the limit $\Gamma_0\gg \Omega$, $\Gamma_0\gg\Gamma$ one arrives
at the following expression for the photo-detection probability:
\begin{equation}
      P(t)=\alpha\frac{\pi S}{r^2}\frac{\hbar\omega_0}{E_{det}}
      \,n(\omega_0)\Gamma_0\frac{\Gamma^2}4
      \bigl[P_1(t)\!+\!2P_{12}(t)\!+\!P_2(t)\bigr]
      \label{2_res}
\end{equation}
where
\begin{eqnarray}
      &&P_{1(2)}(t)=\int\limits_0^\infty\frac{\Gamma_0(t-z)\,
      e^{-\Gamma z}\,d z}{(\delta\Gamma\cos\Omega(t\!-\!z)\pm\frac\Delta2)^2\!+
      \!\frac{\Gamma_0^2(t-z)}4}\\
      &&P_{12}(t)=\int\limits_0^\infty\frac{\Gamma_0(t-z)\cos(\Delta z-\phi)
      \,e^{-\Gamma z}\,d z}{\delta\Gamma^2\cos^2\Omega(t-z)+\frac{\Gamma_0^2(t-z)}4}
\end{eqnarray}
Consider now the extreme limit of infinitely broad spectrum:
$\Gamma_0 \rightarrow\infty$, then in the limit
$\Gamma\gg\Omega$, $\Gamma\ll\Delta$ one has:
\begin{equation}
      P(t)=\alpha\pi\frac{S}{r^2}\,\frac{\hbar\omega_0}{E_{det}}\,
      n(\omega_0)(2\Gamma)\bigl[1-\Omega\tau_g\cos\Omega t].
      \label{d_res}
\end{equation}
Comparing this result with the expression for the photo-detection
probability for the single resonance case~(\ref{weak_b}) one can
see that in the regime of a broad spectrum $\Gamma_0\rightarrow\infty$
the photo-detection probability does not depend on the number of
resonances in FP filter and depends rather on the total
transparency of the filter ($2\Gamma$ in present case).

Consider however the opposite limit of extremely narrow resonances
$\Gamma\ll\Omega$ (this is hardly possible in practice). Then the
resonance factor with respect to separation between resonances,
$\Delta$, appears in the expression for photo-detection
probability:
\begin{eqnarray}
      P(t)\approx\alpha\pi\frac{S}{r^2}\frac{\hbar\omega_0}{E_{det}}
      \,n(\omega_0)(2\Gamma)\left[
      1 -\frac{\Omega\tau_g\Gamma}2\,\frac{\Gamma\cos(\Omega t\!-\!\phi)
      +(\Omega\!-\!\Delta)\sin(\Omega t\!-\!\phi)}{\Gamma^2+(\Omega\!-\!\Delta)^2}
      \right]
      \label{1O_res}
\end{eqnarray}
One can see that there is a parametric resonance for the photo-detection
signal at $\Delta=\Omega$. The photo-detection probability at the
resonance equals to:
\begin{equation}
      P_{\Delta=\Omega}(t)\approx\alpha\frac{\pi S}{r^2}\frac{\hbar\omega_0}{E_{det}}
      \,n(\omega_0)(2\Gamma)\,\bigl[1-\frac{\Omega\tau_g}2\,\cos(\Omega t-\phi)
      \bigr],
\end{equation}
while for $|\Delta-\Omega|\gg\Gamma$ (out of resonance) the term which depends on time in
the photo-detection probability is saturated by the small factor
$\Gamma/|\Delta-\Omega|$.

A similar resonance appears in regime $\Gamma\ll\Omega$ for the case
where the width of the spectrum $n(\omega)$ is broad but $(\omega_0/
\Gamma_0)^2\Omega\tau_g\gg1$ so one can neglect all factors $|1+\Omega
\tau_g\cos\Omega t|$ in eq.~(\ref{2_res}):
\begin{eqnarray}
      P(t)\approx\alpha\frac{\pi S}{r^2}\frac{\hbar\omega_0}{E_{det}}
      \,n(\omega_0)(2\Gamma)\left[
      1-\frac{\delta\Gamma^2}{\Gamma_0^2}\Gamma\,
      \frac{\Gamma\cos(2\Omega t\!-\!\phi)\!+\!(2\Omega\!-\!\Delta)
      \sin(2\Omega t\!-\!
      \phi)}{\Gamma^2+(2\Omega-\Delta)^2}\right]
      \label{2O_res}
\end{eqnarray}
One can see that the parametric resonance at $\Delta=2\Omega$
appears in this case. Now we argue that the resonances at
$\Delta=\Omega$ and $\Delta=2\Omega$~(\ref{1O_res}),
(\ref{2O_res}) arises because of the interference of different
components of the photon's wave packet~(\ref{photon_func}). To
show it let us simply write the transparency of the FP filter as
$t(\omega)=\delta_\Gamma(\omega-\omega_0)+\delta(\omega-\omega_0-\Delta)$
where $\delta_\Gamma(\omega)$ is a narrow peaked function with
finite width $\Gamma\ll\Omega$. Then using the expansion of the
photon wave packet~(\ref{time_fourier}) one can write the
photo-detection probability in terms of the sums over different
components of the photon's wave packet:
\begin{eqnarray}
      P(t)=2\alpha\frac{S}{r^2} \sum\limits_{n,m=-\infty}^\infty
      \int \frac{\hbar\omega}{E_{det}}\,n(\omega)\,d\omega\,
      \,e^{i(n-m)\Omega t}t^*(\omega+n\Omega)
      t(\omega+m\Omega)\,J_{n}(\omega\tau_g) J_m(\omega\tau_g)
\end{eqnarray}
Let us assume that the distance between resonances are exactly equal
to $N\Omega=\Delta$ where $N$ is an integer. Then substituting
the transmission amplitude $t(\omega)$into this expression
and making the integration over $\omega$ one gets:
\begin{eqnarray}
      P(t)\approx4\alpha\frac{S}{r^2}\sum\limits_{n=-\infty}^\infty
      n(\omega_0-n\Omega)\Gamma\,
      \frac{\hbar(\omega_0-n\Omega)}{E_{det}}\,
      \bigl[J_n^2[(\omega_0\!-\!n\Omega)\tau_g]
       +J_n[(\omega_0\!-\!n\Omega)\tau_g]J_{n+N}[(\omega_0\!-\!n\Omega)\tau_g]
      \cos N\Omega t\bigr]
\end{eqnarray}
Then for the special choice $N=2$ assuming that the argument of the Bessel
function is typically large one can put $J_{n+2}(x)\approx -J_n(x)$ for
$x\gg1$. Then the photo-detection probability can be written as
\begin{eqnarray}
      P(t)\approx4\alpha\frac{S}{r^2}\,\bigl(1-\cos2\Omega t\bigr)
      \sum\limits_{n=-\infty}^\infty
      n(\omega_0-n\Omega)\Gamma\,\frac{\hbar(\omega_0-n\Omega)}{E_{det}}
      \,J_n^2[(\omega_0-n\Omega)\tau_g].
\end{eqnarray}
One can see that the term which is alternating in time in $P(t)$ arises
due to interference of the different components of the photon wave
packet~(\ref{photon_func}).

\subsection{Analysis of intensity correlations}

In the previous section, we showed that by appropriate filtering
of the light signal, it is possible to extract oscillations
associated with the past interaction with the GW. While this
previous proposal is promising and attractive, one can take an
alternative route for the detection of the GW signal: the measurement
of intensity--intensity correlations of the photo-detection signal.
Consider the correlator $\langle \langle I(t_1)
I(t_2)\rangle\rangle$: because of the time modulation due to the
presence of the GW, it no longer depends on the variable $t_1-t_2$
only. Using the definition the intensity operator,  one can see
qualitatively that the Fourier components of  Eq.
(\ref{time_fourier}) will lead to a two particle interference
effect.

The quantity to be measured on the Earth is thus the intensity correlator
of the photo-detection signal which is the time ordered product of
electrical field operators:
\begin{equation}
      \langle\langle:\hat I(t_1)\hat I(t_2):\rangle\rangle\,dt_1 dt_2
      =\alpha^2\frac{S^2c^2\,dt_1\,dt_2}{E_{det}^2}
      \,\langle\langle{\bf\hat E}^{(-)}(t_1){\bf\hat E}^{(-)}(t_2)
      {\bf\hat E}^{(+)}(t_2){\bf\hat E}^{(+)}(t_1)\rangle\rangle
\end{equation}
Substituting here the expression for the electrical field operators~(\ref{pos_e_op})
and performing the quantum average one gets
\begin{eqnarray}
      &&\langle \langle:\hat I(t_1)\hat I(t_2):\rangle\rangle\,dt_1 dt_2=
      2\alpha^2 \frac{S^2\,dt_1dt_2}{r^4}\int d\omega_1d\omega_2\,
      e^{-i\omega_2[t_1-t_2+\tau_g(\sin\Omega t_1-\sin\Omega t_2)]}
      \nonumber\\
      &&\qquad\qquad\qquad
      \times\,\frac{\hbar^2\omega_1
      \omega_2}{E_{det}^2}\,n(\omega_1)n(\omega_2)\,e^{i\omega_1[t_1-t_2+\tau_g(\sin\Omega t_1-
      \sin\Omega t_2)]}.
\end{eqnarray}
This correlator can be rewritten as
\begin{equation}
      \langle \langle:\hat I(t_1)\hat I(t_2):\rangle\rangle\,
      dt_1dt_2=2\alpha^2\frac{S^2\,dt_1dt_2}{r^4}\,
      \bigl|f[t_1-t_2+\tau_g(\sin\Omega t_1-
      \sin\Omega t_2)]\bigr|^2,
\end{equation}
where
\begin{equation}
      f(t)=\int \frac{\hbar\omega}{E_{det}}\,n(\omega)\,e^{-i\omega t}\,d\omega~.
\end{equation}
The function $f(t)$ typically decays exponentially on a time scale
given by the inverse of the line width $\Gamma_0\gg\Omega$. Thus
one can safely linearize the dependence of the intensity-intensity
correlator with respect to the relative time $\tau=t_1-t_2$:
\begin{eqnarray}
      \langle \langle:\!\hat I(t+\tau)\hat I(t)\!:\rangle\rangle=
      2\alpha^2\frac{S^2}{r^4}\bigl|f[\tau(1+\Omega\tau_g\cos\Omega t)]\bigr|^2
\end{eqnarray}
where $\tau=t_1-t_2$ is the relative time and $t$ is the total time.
Let us now calculate the spectral power, $S(t)$ of the photons' noise
defined as
\begin{equation}
      S(t)=\int \langle \langle:\!\hat I(t+\tau)\hat I(t)\!:\rangle\rangle
      \,d\tau
\end{equation}
Substituting here the expression for photon intensity correlator one
can show that
\begin{equation}
       S(t)=\frac{4\pi\alpha^2}{1+\Omega\tau_g\cos\Omega t}
       \,\frac{S^2}{r^4}\int
       \frac{(\hbar\omega)^2}{E_{det}^2}\,n^2(\omega)\,d\omega
\end{equation}
One can see from this equation that the intensity correlation
contain periodic oscillations -- although with a small amplitude
-- in the total time. Note that contrary to the Fabry-Perot
diagnosis, here a rather accurate measurement is implied, as the
contribution to other noise sources (due to scattering...) needs
to be minimized compared to this periodic signal. In particular,
it requires a large time acquisition window to filter out spurious
fluctuations.

\subsection{Conclusion}

Let us now summarize all results obtained through the article and
give some estimations and limitations for the proposed
gravitational waves detection method. The primary effect we have
dealt with is the modulation of the photon phase due to
interaction with gravitational field have been found in the
geometrical optics approximation and described by
eq.~(\ref{photon_func}) for the photon wave function:
\begin{equation}
      \phi_k(\xi)={\cal C}(\xi)\,e^{-i\omega_k\xi-
      i\omega_k\tau_g\sin\Omega\xi},
\end{equation}
where $\xi=t-r/c$ and ${\cal
C}(\xi)=\bigl[1+\Omega\tau_g\cos\Omega\xi\bigr]^{-1}$ is the
pre-exponential factor. This effect of phase modulation leads
immediately to well known Shapiro effect of the time delay of the
signal propagating near the source of gravitational field. For the
case of the source of gravitational waves this time delay is an
alternating function of time $\tau_{ar}=r/c-\tau_g\sin\Omega\tau$,
where $\tau_{ar}$ is an arrival time of the signal, $r/c$ the
traveling time of the signal in flat Minkovski space and
\begin{equation}
      \tau_g=-\frac{h_{xx}(\Lambda)}\pi\,\left(\frac{\Lambda}{D}\right)^2\,
      \Omega^{-1}.
\end{equation}
Here $h_{xx}(\Lambda)$ is the dimensionless strength of
gravitational wave near the GW source, $D$ impact parameter,
$\Omega=2\pi c/\Lambda$ is the frequency of gravitational waves.
The Shapiro effect in fact seems to be a very promising candidate for
gravity wave detection experiment since the time delay (or the
additional alternating in time phase) mainly accumulated near the
source of gravitational waves where the strength of GW many order
of magnitude greater when they reach the Earth. However in the
proposed experiment~\cite{sazhin_1} it requires a pulsar on the
line of sight which passes near the source of gravitational
radiation. In this paper we propose a gravitational
waves detection method based on the Fabry-Perot interference
analysis (or equivalently time correlation measurement) of the
light signal from arbitrarily light source (not necessarily a
pulsar) passing near the strong source of gravitational radiation.
Then the photodetector signal contains an alternating in time
component (with GW frequency) proportional to the strength of
gravitational radiation near the the GW source (see
eq.~\ref{weak_b}, \ref{d_res}):
\begin{equation}
      P(t)= P_0(1-\Omega\tau_g\cos\Omega t),
\end{equation}
with $P_0$ a photodetector signal corresponding to a light coming
to a FP filter from distant star. To estimate an effectiveness of
the proposed method let us first estimate a brightness of the light
source needed to resolve an alternating component in the
photo-detection signal. Collecting the signal over long time
$\tau_{\rm obs}$ allow to better resolve the alternating
component. Assuming that during the observation time $N_{\rm obs}$
photons coming from a distant star hit the photodetector, in the
limit of Poissonian statistics, $\langle \delta N_{\rm
obs}^2\rangle=N_{\rm obs}$ one has the following limiting
requirement:
\begin{equation}
      \Omega\tau_g(D)\,\sqrt{N_{\rm obs}}\gg 1.
\end{equation}
in the extreme case where the light passes in immediate proximity
($D\approx\Lambda$) to the GW source one has the condition:
\begin{equation}
      h(\Lambda)\sqrt{N_{\rm obs}}\gg 1.
\end{equation}
Then assuming that the distant star emits in the frequency band
$\Delta\omega$ and that the diameter of the telescope equals to $d$ one
has:
\begin{equation}
      N_{\rm obs}\sim L\,\frac{d^2}{R^2}\,
      \frac{\Gamma}{\Delta\omega}\,\frac{\tau_{\rm obs}}{\hbar\omega_0},
      \label{br}
\end{equation}
where $L$ is the brightness of the distant star, $R$ is the
distance to the Earth.

The most promising candidates of gravitational waves sources
appropriate for proposed detection method are the periodic sources
of gravitational radiation such as close binaries at the final
stage of evolution or a asymmetric rotating neutron stars,
allowing to collect the signal long enough time to resolve the
tiny gravitational wave modulation on the stochastic Poisson
noise. Let us give some estimations, assuming the total mass, $m$
of binary system of the order of the mass of the Sun, and rotating
with a frequency $\nu\sim 10^{-2}$ Hz. Then the dimensionless
strength of gravitational waves at the distance $r$ from the
emitter, according to quadrupole approximation, is given by
\begin{equation}
      h\approx 4\frac{G^{5/3}}{c^4}\,\frac{1}{r}\,m^{5/3}\nu^{2/3}
\end{equation}
assuming that the role of the distant star plays one of the
components of the binary system $r=(Gm/\nu^2)^{1/3}$ one has
$h(r)\approx4(Gm\nu)^{4/3}/c^4\sim 10^{-9}$. In order to resolve
this modulation one needs about $N_{\rm obs}\sim 10^{18}$ photons
coming to a photodetector during $\tau_{\rm obs}$. Assuming the
the typical distance from the binary system to the Earth of the order
$1$ kpa  with $d\approx 10\;m$, $\omega_0\sim 10^{14}$ Hz and
$\Gamma/\Delta\omega\sim 1$, $\tau_{\rm obs} \sim 10^3\nu^{-1}\sim
10^5 s$ one has the required brightness of the binary component,
according to eq.~(\ref{br})of the order $L\sim 10^5 L_\mathrm{\rm
S}$ where $L_\mathrm{\rm S}$ is the brightness of the Sun. Of
course the requirement to have such a bright component of the binary
system makes the observation of the proposed effect rather
problematic in practice.

In reality the situation with binary systems is even worse due to
the presence of the Doppler effect for the light emitted from one
of the rotating components. In fact the Doppler effect results in the
same frequency modulation of the emitted light as the
gravitational waves modulation and the amplitude of the frequency
modulation due to the Doppler effect is of the order of
\begin{equation}
      \left(\frac{\Delta\omega}\omega\right)_\mathrm{Doppler}\sim
      \left(\frac{v}{c}\right)^2\approx\frac{(Gm\nu)^{1/3}}c
      \,\frac{\nu}{c}.
\end{equation}
For parameters described above it results in an effect of the order
$10^{5}$ that is five orders of magnitude greater than the frequency
modulation due to gravitational waves.

The situation however could be much better for very close neutron
star binaries at the last stage of evolution. In this case the
rotational frequency can be of the order of $10^2$ Hz and results
in a dimensionless strength of emitted gravitational waves of the
order $h(\Lambda)\sim 10^{-6}$. Assuming that the distance to the
binary neutron star system is of the order $1$ kpa  and that there
is a bright distant star situated on the line of sight to the
binary from the Earth with impact parameter $D\sim \Lambda$ and
collecting the signal during the time $\tau_{\rm obs}\sim
10^3\nu^{-1}\sim 10\; s$ one has the following lower limit for the
brightness of distant star $L\sim 10^{2} L_\mathrm{S}$.

\subsection{Acknowledgments.}

We acknowledge discussions to M. Sazhin, A. Starobinsky and K.
Bayandin. A. L. thanks the Ecole-Normale Landau Institute
agreements and CNRS for his stay at the Centre de Physique
Th\'eorique. He also acknowledges the Forschungszentrum Juelich
for financial support within the framework of the Landau Program.
We acknowledge financial support from the Russian Science Support
Foundation, the Russian Ministry of Science, and the program
`Quantum Macrophysics' of the RAS.


\end{document}